\documentclass[aps,prl,reprint,groupedaddress,longbibliography ]{revtex4-1}
\usepackage{epsf}
\usepackage{epsfig}
\usepackage{graphicx}
\usepackage{color}
\usepackage{siunitx}
\usepackage{epstopdf}
\usepackage{amsmath}
\usepackage{braket}
\graphicspath{ {./figures/} }
\begin{document}
\newcount\colveccount
\newcommand*\colvec[1]{
        \global\colveccount#1
        \begin{pmatrix}
        \colvecnext
}
\def\colvecnext#1{
        #1
        \global\advance\colveccount-1
        \ifnum\colveccount>0
                \\
                \expandafter\colvecnext
        \else
                \end{pmatrix}
        \fi
}
\title {\large
The electric non-dipole effect in strong-field ionization
}
\author{A. Hartung$^{1}$}
\author{S. Brennecke$^{2}$}
\author{K. Lin$^{1,3}$}
\email{lin@atom.uni-frankfurt.de}
\author{D. Trabert$^{1}$}
\author{K. Fehre$^{1}$}
\author{J. Rist$^{1}$}
\author{M. S. Sch\"offler$^{1}$}
\author{T. Jahnke$^{1}$}
\author{L. Ph. H. Schmidt$^{1}$}
\author{M. Kunitski$^{1}$}
\author{M. Lein$^{2}$}
\author{R. D\"{o}rner$^{1}$}
\email{doerner@atom.uni-frankfurt.de}
\author{S. Eckart$^{1}$}
\email{eckart@atom.uni-frankfurt.de}
\affiliation{$^1$ Institut f\"ur Kernphysik, Goethe-Universit\"at, Max-von-Laue-Str. 1, 60438 Frankfurt, Germany \\
$^2$ Institut f\"ur Theoretische Physik, Leibniz Universit\"at Hannover, Appelstr. 2, 30167 Hannover, Germany \\
$^3$ State Key Laboratory of Precision Spectroscopy, East China Normal University, 200241, Shanghai, China}
\date{\today}
\begin{abstract}
Strong-field ionization of atoms by circularly polarized femtosecond laser pulses produces a donut-shaped electron momentum distribution. Within the dipole approximation this distribution is symmetric with respect to the polarization plane. The magnetic component of the light field is known to shift this distribution forward. Here, we show that this {\em magnetic} non-dipole effect is not the only non-dipole effect in strong-field ionization. We find that an {\em electric} non-dipole effect arises that is due to the position dependence of the electric field and which can be understood in analogy to the Doppler effect. This {\em electric} non-dipole effect manifests as an increase of the radius of the donut-shaped photoelectron momentum distribution for forward-directed momenta and as a decrease of this radius for backwards-directed electrons. We present experimental data showing this fingerprint of the {\em electric} non-dipole effect and compare our findings with a classical model and quantum calculations.
\end{abstract}
\maketitle

The ionization of an atom by the interaction with an electromagnetic wave is often described by only considering the temporal evolution the electric field vector. This is at the heart of the dipole approximation which neglects the magnetic component and the position-dependence of the light field. It is a surprisingly good approximation over a wide range of wavelengths and intensities from the perturbative single-photon ionization regime of the photoelectric effect over multiphoton ionization to strong-field non-relativistic tunnel ionization. Single-photon ionization is dominated by electric dipole transitions \cite{Krause1969,Cooper1993}. The same holds in the non-relativistic strong-field regime where it is the time-dependent {\em electric} field that drives tunnel ionization and determines the electron momentum distribution in a good approximation \cite{Keldysh1965,Klaiber2005}. Within the dipole approximation, the momentum distribution of the emitted electrons is forward-backward symmetric for single-photon ionization \cite{Grundmann2018} as well as for strong-field ionization \cite{Misha2005}. 

\begin{figure*}
\epsfig{file=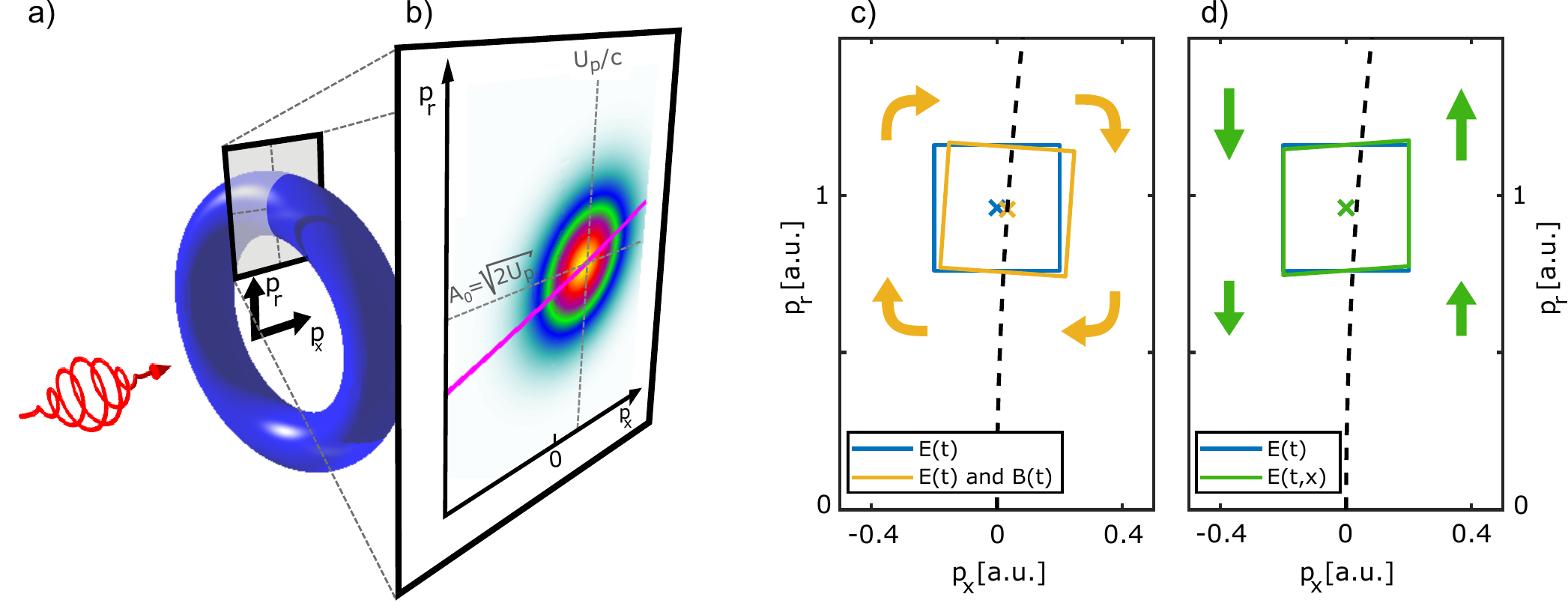, width=17.9cm}
\caption{(a) Illustration of the donut shape of the electron momentum distribution from strong-field ionization in a circularly polarized laser pulse and of the coordinate system that is used throughout our paper. (b) Artist's view of a combination of the two non-dipole effects. The {\em magnetic} component of the light field drives the donut forward by about $U_p/c$ through the magnetic part of the Lorentz force. The temporospatial {\em electric} field results in an increase of $\braket{{p}_r}$ as a function of $p_x$ (purple line in (b)). (c) Four classical trajectories are calculated using the dipole approximation. The initial momenta after tunneling are chosen such that the final electron momenta mark the edges of the blue square. Using the same initial conditions but including the magnetic component only (no spatial dependence of the electric field) leads to the yellow quadrangle in (c) which is forward shifted and rotated with respect to the blue square. For the green quadrangle in (d) the temporospatial electric field is considered and the magnetic field is neglected which leads to shearing of the electron momentum distribution as indicated. In (c) and (d) the crosses indicate the centers of the quadrangles and the dashed lines correspond to the naively expected forward shift of $0.5 p_r^2/c$. For visual representation, both effects are amplified by a factor of 10 (see text).} 
\label{fig_figure1label} 
\end{figure*}

The leading physical mechanisms which eventually lead to a failure of the dipole approximation and a breaking of the forward-backward symmetry of electron emission are different for single-photon and strong-field ionization. For single-photon ionization the dominating term beyond electric dipole transitions is due to electric quadrupole transitions. The interference between electric dipole and electric quadrupole transitions leads to a breaking of the symmetry and at high photon energies to a forward emission of photoelectrons and backward emission of ions \cite{Sommerfeld1930,Cooper1993,Seaton1995,Forre2014,grundmann2020}. The electric quadrupole transitions are due to the spatial dependence of the electric field whereas the transition amplitudes that are driven by the magnetic component of the field are typically much weaker \cite{Cooper1993}. In contrast, for strong-field ionization the breakdown of the dipole approximation \cite{Klaiber2005,Smeenk2011_B,He2017,Chelkowski2018,Brennecke2018,Haram2020} is commonly argued to be due to the magnetic component of the light field that drives the electron forward via the Lorentz force \cite{Katsouleas1993} (for the effect of rescattering see Refs. \cite{Walser2000,Dammasch2001,Palaniyappan2005,Emmanouilidou2017}). However, the temporospatial dependence of the electric field has not been discussed so far for strong-field ionization at non-relativistic intensities. Up to now, no experimental evidence has been presented which shows an influence of the temporospatial nature of the light field in strong-field ionization that breaks the forward-backward symmetry. It is the purpose of the present paper to provide that missing experimental evidence and show how the temporospatial dependence of the electric field alters electron momentum distributions in strong-field ionization. We show that electric field driven non-dipole effects are as important as those caused by the magnetic field, if a suitable observable is looked at.

We consider circularly polarized light, to avoid the additional complexity caused by recollisions of the electron with its parent ion. Upon ionization of an atom by a circularly polarized, multi-cycle laser pulse at a wavelength of 800\,nm the electron momentum distribution has a donut-like shape (blue shape in Fig. \ref{fig_figure1label}(a)). The radius of the donut in momentum space is approximately given by 
\begin{equation}
\braket{{p}_r}=A_0=\frac{E}{\omega} \, ,
\label{simpleEq}
\end{equation}
where $p_r=\sqrt{p_y^2+p_z^2}$ is the radial momentum component perpendicular to the light propagation direction, $E$ is the laser's peak electric field, $\omega$ is the central frequency of the laser pulse and $A_0$ is the laser's peak vector potential (atomic units are used unless stated otherwise). When calculated within the dipole approximation, this donut is symmetric with respect to the polarization plane ($p_yp_z$-plane). In pioneering work Smeenk at al. showed experimentally that this momentum distribution is slightly forward shifted by about $U_p/c$ \cite{Smeenk2011_B} ($U_p=\frac{E^2}{2\omega^2}$ is the ponderomotive energy and $U_p/c$ is the forward momentum which classical mechanics predicts for an electron launched with zero initial velocity and accelerated by a circularly polarized  {\em spatially homogeneous} electromagnetic field.) Soon after, Klaiber et al. \cite{klaiber2013} predicted an additional forward shift of $I_p/(3c)$ (with the ionization potential $I_p$) which was confirmed by other calculations \cite{Chelkowski2014,Chelkowski2015,He2017} and an experiment \cite{hartung2019magnetic}. In all cases, these forward shifts were discussed to be due to the magnetic component of the light field.

Fig. \ref{fig_figure1label}(b) shows an artist's view of the full non-dipole effect in strong-field ionization (including the magnetic and the electric non-dipole effect) in cylindrical coordinates ($p_x$, $p_r$). To further investigate these non-dipole effects, we use classical trajectory simulations (CTS) consisting of two steps. In the first step the electron is freed by laser-induced tunnel ionization. In a second step the electron's acceleration in the time- and position-dependent electromagnetic field is described classically by Newton's equation. For this second step the Coulomb interaction of the electron and its parent ion are not taken into account \footnote{CTS with Coulomb interaction are sensitive to the choice of the tunnel exit position \cite{Ni2018_theo}. CTS using tunnel exit positions that take non-adiabaticity and non-dipole effects into account are beyond the scope of the current work.}. The CTS model is well-suited to distinguish electric and magnetic field driven non-dipole effects. We assume adiabatic tunneling, i.e. that after tunneling the electrons have zero initial velocity in tunnel direction and their velocity in both directions perpendicular to the tunnel direction is described by a rotationally-symmetric initial Gaussian momentum distribution that is centered at zero momentum (analogous to Eq. (9) of Ref. \cite{Shilovski2016}). The electric and magnetic field of the circularly polarized laser pulse are defined by:
\begin{eqnarray}  
\begin{aligned} 
\vec{E}(x,t) &=  E_0(t)   \colvec{3}{0}{\cos{(\omega t - \zeta \frac{\omega}{c} x)}} {\sin{(\omega t - \zeta \frac{\omega}{c} x)}} ,\\
\vec{B}(x,t) &=  \chi \frac{E_0(t)}{c}   \colvec{3}{0}{-\sin{(\omega t - \zeta \frac{\omega}{c} x)}} {\cos{(\omega t - \zeta \frac{\omega}{c} x)}} .
\end{aligned}
\label{eqn1}
\end{eqnarray}   
Here, $E_0(t)$ is the temporal envelope of the light pulse. For a real light pulse one sets $\zeta=\chi=1$ while in the dipole approximation one sets $\zeta=\chi=0$. 

Taking only the magnetic field into account ($\zeta=0\mathrm{,\,}\chi=1$) leads to the well-known forward shift of the donut-shaped electron momentum distribution by $U_p/c$ \cite{Smeenk2011_B} and an additional internal rotation of the electron momentum distribution around its center by an angle of about $A_0/c$ as schematically illustrated in Fig. \ref{fig_figure1label}(c). Importantly, the most probable radial momentum $\braket{{p}_r}$ of the distribution as a function of $p_x$ remains constant in this case. 

In full analogy, the effect of the temporospatial dependence of the electric field $\vec{E}(x,t)$ alone ($\zeta=1\mathrm{,\,}\chi=0$) can be included in the calculations. The temporospatial electric field $\vec{E}(x,t)$ leads to a shearing of the final momentum distribution as compared to an electric field $\vec{E}(t)$ that is only time-dependent (see Fig. \ref{fig_figure1label}(d)). 

Strikingly, the increase of the radius of the donut-shaped electron momentum distribution as a function of $p_x$, which is shown in Fig. \ref{fig_figure1label}(b) from an artist's perspective, can be interpreted in analogy to the Doppler effect. In the lab frame, electrons that propagate parallel [anti-parallel] to the light-propagation direction experience an oscillating force with a frequency that is lower [higher] than the central frequency of the incident laser field. The frequency of this oscillating force is given by $\bar{\omega}(p_x)=\omega (1-p_x/c)$ for a light propagation direction that is parallel to $p_x$. Using this insight one can approximate the most probable radial electron momentum $\braket{{p}_r}$ for a given value of $p_x$. To this end Eq. \ref{simpleEq} is generalized  using $\bar{\omega}(p_x)$ instead of $\omega$ which leads to
\begin{eqnarray}
\begin{aligned}
\braket{{p}_r}(p_x)&= \frac{E}{\bar{\omega}(p_x)}\approx\left(1+\alpha\, \frac{p_x}{c}\right) A_0
\end{aligned}
\label{effpr}
\end{eqnarray}
where $\alpha=1$ in this simplest case. Later, we will use $\alpha$ as a fitting parameter to analyze the {\em electric} non-dipole effect in experimental data and compare the results to more sophisticated theoretical models. But in a first step, the parameter $\alpha$ is extracted from our CTS simulations for several scenarios that are summarized in Tab. 1. It is evident that $\alpha\approx 1$ if the temporospatial dependence of the electric field is taken into account (scenarios V and W in Tab. 1). In particular, the magnetic field does not significantly change the value of $\alpha$.

\begin{table}
	\centering
	\begin{tabular}[t]{c|rl|c|c|c}
		scenario & \multicolumn{2}{|c|}{light field's definition}& $\zeta$ &$\chi$ & $\alpha$ \\
		\hline
		T & \hspace{2.3em}$\vec{E}(t)$,& no magnetic field &$0$& $0$ & $0.00$\\
		U & $\vec{E}(t)$,& $\vec{B}(t)$  &$0$& $1$ & $0.00$\\
		V & $\vec{E}(\vec{x},t)$,& no magnetic field  &$1$& $0$ & $1.07$\\
		W & $\vec{E}(\vec{x},t)$,& $\vec{B}(t)$  &$1$& $1$ & $1.07$				
	\end{tabular}
	\caption{The parameter $\alpha$ at $p_x = 0$ \,a.u. is determined from the CTS model for various scenarios regarding the definition of the electromagnetic field (Eq. \eqref{eqn1}) using a $\sin^2$-envelope with a total duration of  12 cycles.}
\end{table}

The shearing of the momentum distribution which is induced by the {\em electric} non-dipole effect is quantified by the value of $\alpha$. The key point of the present paper is to show this shearing in an experiment and in numerical \textit{ab-initio} simulations of the TDSE (time-dependent Schrödinger equation).

\begin{figure}
	\epsfig{file=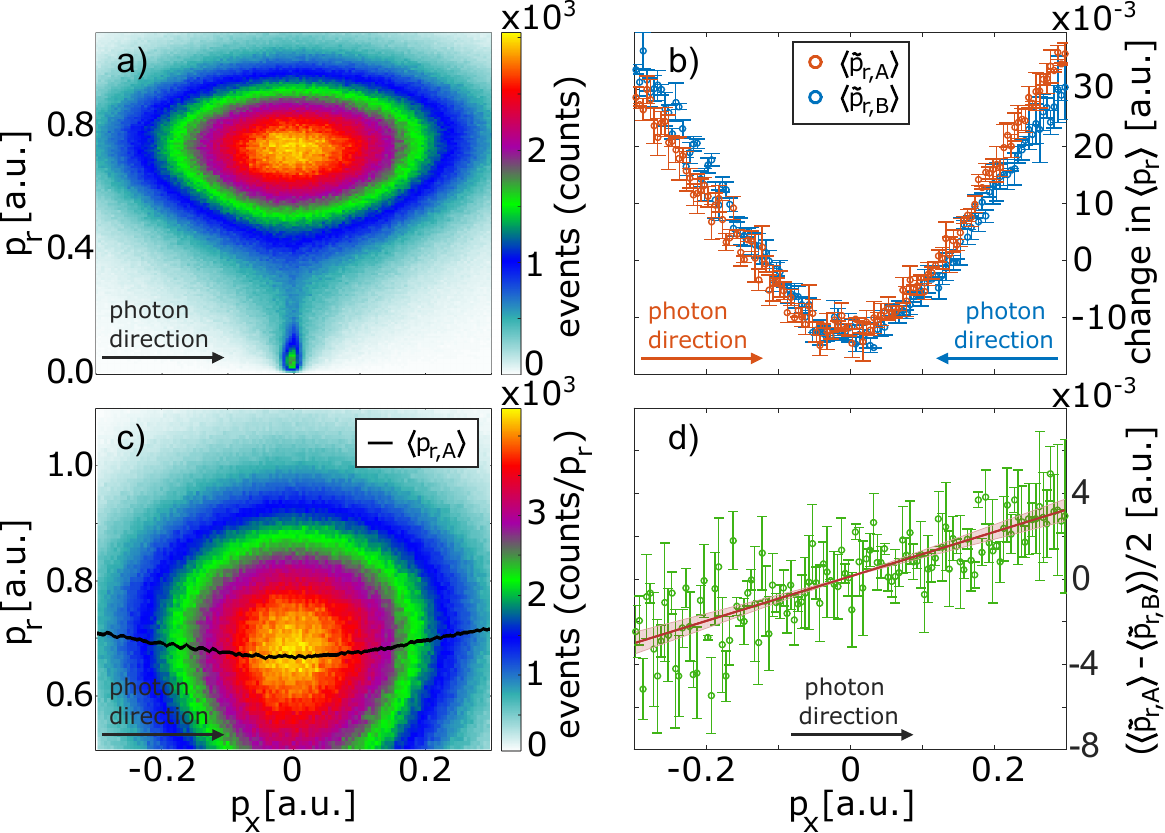, width=8.7cm}
	\caption{Experiment on the strong-field ionization of xenon by circularly polarized light at a wavelength of 800 nm and an intensity of $8.1\cdot10^{13}$\,W/cm$^2$. (a) shows the electron momentum resolved on the momentum along the light propagation-direction, $p_x$, and the radial momentum in the plane of polarization, $p_r=\sqrt{p_y^2+p_z^2}$, for a laser beam that has a propagation-direction as indicated ($k_{\mathrm{ph}}>0$). (c) shows the same as (a) but zooming in and showing the most probable radial momentum, $\braket{p_{r,A}}$, as a function of $p_x$ (black line). The red data points $\braket{\tilde{p}_{r,A}}$ in (b) show the same as the black line in (c) after subtracting a constant value (see text). The blue data points, $\braket{\tilde{p}_{r,B}}$, show the analogue to the red data points but for an inverted light propagation-direction ($k_{\mathrm{ph}}<0$). (d) $S(p_x)=\frac{\braket{\tilde{p}_{r,A}}(p_x)-\braket{\tilde{p}_{r,B}}(p_x)}{2}$ is shown in green. The red line is a linear fit to the green data points using $\alpha$ as a free parameter for $S(p_x)=\alpha\frac{p_x}{c}A_0$. The value of $\alpha A_0/\mathrm{a.u.}=1.43\pm 0.23$ is obtained from the linear fit and is used as a figure of merit to describe the {\em electric} non-dipole effect. The error bars show the standard deviation of the statistical errors.}
	\label{fig_figure2label} 
\end{figure}

The experiment was performed using the same specialized COLTRIMS reaction microscope \cite{ullrich2003recoil} and the same laser setup as in Ref. \cite{hartung2019magnetic} (also see Supplementary Material \cite{SupplementaryMaterial}). The 25-fs laser pulses with a central wavelength of 800 nm and a repetition rate of 10 kHz are split into two counter-propagating pathways A and B. In pathway A [B] the light propagation-direction is parallel [anti-parallel] to $p_x$ and thus the light's wavevector $k_{\mathrm{ph}}$ is positive [negative]. In both pathways lambda-quarter and lambda-half waveplates ensure that the laser pulses that enter the vacuum chamber are circularly polarized. The two pulses are focused from opposite sides to the same spot in the xenon-gas jet. Shutters in pathways A and B toggled between using the two possible pathways. This procedure allows us to eliminate most systematical errors which is essential, since the expected changes of the radius of the donut are on the order of 0.001\,a.u. 

Figure \ref{fig_figure2label}(a) shows the measured electron momentum distribution in cylindrical coordinates after integration over the angle in the polarization plane (see Figs. \ref{fig_figure1label}(a) and \ref{fig_figure1label}(b)). Here,  the intensity of the laser pulses in the focus is $8.1\cdot 10^{13}$\,W/cm$^2$ (peak electric field of 0.034\,a.u.). Fig. \ref{fig_figure2label}(c) shows the same data as Fig. \ref{fig_figure2label}(a) after restricting the momenta to $0.5$\,a.u.$<p_r<1.1$\,a.u. The forward shift of the electrons with $0.5$\,a.u.\,$<p_r<1.1$\,a.u. that is due to the magnetic field is not visible in Fig. \ref{fig_figure2label}(c) because it is only $\braket{p_x}=0.0021 \pm 0.0001$\,a.u. (value has been determined by a Gaussian fit). This forward shift is due to the magnetic field and can be compared with the theoretically expected value $\braket{p_x}\approx p_r^2/(2c) + I_p/(3c) =0.0027$\,a.u. for $p_r=0.68$\,a.u. \cite{hartung2019magnetic}. (The experimental uncertainty of $\braket{p_x}$ only takes the statistical error into account.) 

From the data shown in Fig. \ref{fig_figure2label}(c), we have determined the most probable radial momentum using a Gaussian fit for every bin along $p_x$. The maximum of these Gaussian fits is shown by the black line and referred to as $\braket{p_{r,A}}$. The values of $\braket{p_{r,A}}$ as a function of $p_x$ have a close to parabolic shape. In a next step the light propagation-direction is inverted in the experiment and the analysis is done again and the result is referred to as  $\braket{p_{r,B}}$. For each of the two resulting close to parabolic shapes the mean is subtracted using $\braket{\tilde{p}_{r,A}}(p_x)=\braket{p_{r,A}}(p_x)-q_A$ and $\braket{\tilde{p}_{r,B}}(p_x)=\braket{p_{r,B}}(p_x)-q_B$. Here, the scalar value $q_A$ is the mean in $p_r$ obtained from a Gaussian fit using the projection of the data shown in Fig. \ref{fig_figure2label}(c). $q_B$ is the mean for pathway B in full analogy.

The results for $\braket{\tilde{p}_{r,A}}$ and $\braket{\tilde{p}_{r,B}}$ are presented in Fig. \ref{fig_figure2label}(b). Within the dipole approximation the close to parabolic shape would be forward-backward symmetric and the $p_x$ dependence would be caused by non-adiabatic effects and Coulomb interaction of the electron with its parent ion after tunneling \cite{Eckart2018_Offsets}. To disentangle the symmetric contributions from the non-dipole effects, the difference 
\begin{equation}
S(p_x)=\frac{\braket{\tilde{p}_{r,A}}(p_x)-\braket{\tilde{p}_{r,B}}(p_x)}{2}
\end{equation}
is shown in Fig. \ref{fig_figure2label}(d). The slope of $S(p_x)=\alpha A_0 p_x/c$ is found by fitting and we obtain a value of $\alpha=2.38\pm 0.38$ (which includes the statistical error only). We estimate the systematic error of $\alpha$ to be $\pm0.42$. As suggested by the illustration in Fig. \ref{fig_figure1label}(b), we find that $\braket{p_r}(p_x)$ linearly increases as a function of $p_x$ for $k_{\mathrm{ph}}>0$. Thus, the electrons flying in the forward direction show a larger radial momentum than those that are emitted into the backward direction. So qualitatively, the experimental findings are in line with the expectation from Eq. \ref{effpr}. We have repeated the experiment \cite{SupplementaryMaterial} using an intensity of $6.8\cdot10^{13}$\,W/cm$^2$ ($1.2\cdot10^{14}$\,W/cm$^2$) and obtained $\alpha =1.51 \pm 0.31$ ($\alpha\ =3.10\ \pm\ 0.10$)  \footnote{Analysis of another data set for the strong field ionization of argon using circularly polarized light at 800\,nm yields a value of $\alpha=(1.315\pm 0.485)/(A_0/\mathrm{a.u.})=1.69\pm 0.62$ using $A_0=0.78$\,a.u. (see Fig. 6.57 in Ref. \cite{HartungDiss}).}. Thus, we find that the experimentally obtained value of $\alpha$ increases as a function of the light intensity \footnote{In order to judge if saturation is relevant in the experiment, we have analyzed the changes of $\braket{{p}_r}$ as a function of the laser intensity. We found that for our 25-fs laser pulses with a central wavelength of 800 nm the single ionization of xenon is saturated [not saturated] for an intensity of $1.2\cdot10^{14}$\,W/cm$^2$ [$6.8\cdot10^{13}$\,W/cm$^2$]. This is consistent with Ref. \cite{dimauro1995ionization} where the saturation intensity of xenon was found to be $7\cdot10^{13}$\,W/cm$^2$ using 100-fs laser pulses with a central wavelength of 620 nm.}.

\begin{figure}
\epsfig{file=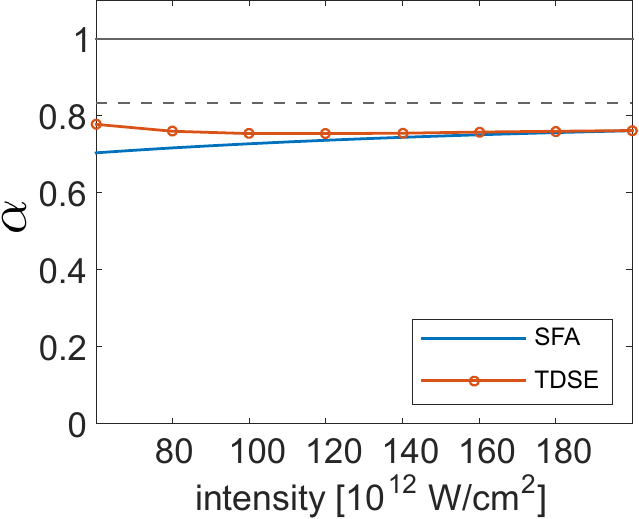, width=4.6cm}
\caption{Dependence of $\alpha$ at $p_x = 0$ \,a.u. on the intensity for a xenon atom at a wavelength of 800\,nm that is obtained from a numerical solution of the TDSE (red line) and using the SFA (blue line). The adiabatic limit of $\alpha=5/6$ (gray thick dotted line) and the estimate based on Eq. \ref{effpr} (gray thick solid line) are shown as horizontal lines.}
\label{fig_figure3label} 
\end{figure}

For a quantitative comparison we have performed numerical simulations of the 3D TDSE including non-dipole effects to first order in $1/c$ \cite{Brennecke2018,hartung2019magnetic} and using an effective potential for xenon \cite{Tong2005} converted into a pseudopotential for the 5p state at cutoff radius $r_{cl}=2$~a.u. \cite{Troullier1991}. Based on the corresponding momentum distribution for a short laser pulse with $\sin^2$-envelope of 3 cycles total duration, the value of $\alpha$ is obtained by a linear fit to $S_{\mathrm{TDSE}}(p_x)=\frac{\braket{p_{r}}(p_x)-\braket{p_{r}}(-p_x)}{2}$. The obtained values of $\alpha$ are shown in Fig. \ref{fig_figure3label} for a wide range of intensities. The values of $\alpha$ are in the range from $0.75$ to $0.78$ and are systematically smaller than the naively expected value of $\alpha\approx 1$ (see Tab. 1 and Eq. \ref{effpr}).

To deepen our understanding of the electric non-dipole effect theoretically, we have also studied a quantum-orbit model derived from the strong-field approximation \cite{Walser2000,He2017} (SFA) by application of a saddle-point approximation (analogous to the procedure in Ref. \cite{hartung2019magnetic}). The resulting values of $\alpha$ are in good agreement with the TDSE result (see Fig. \ref{fig_figure3label}) \footnote{In principle, more general approaches could be considered for the quantum orbit model \cite{PhysRevA.63.011403}.}. In particular, this shows that the long-range ionic potential, which is not included in SFA, does not significantly influence the value of $\alpha$. The most important difference comparing the SFA to the CTS model is that the SFA incorporates initial momentum offsets, i.e. the initial distribution of the freed electrons is not rotationally-symmetric at the tunnel exit \cite{Eckart2018_Offsets}. In first order of $1/c$ and leading order of the Keldysh parameter $\gamma=\sqrt{2 I_p}/A_0$ the most probable momentum as a function of $p_x$ is given by 
\begin{equation}
\braket{{p}_r}(p_x)=A_0+\frac{1}{3A_0}\left(I_p+\frac{p_x'^2}{2}\right)+\frac{A_0}{c}p_x,
\label{Eq4}
\end{equation}
with the shifted momentum $p_x'=p_x-(U_p+2I_p/3)/c$ (for a light propagation-direction that is parallel to $p_x$). The second quadratic term that is due to non-adiabatic offset momenta is centered around the global maximum of the momentum distribution. Using Eq. \ref{Eq4} it can be shown, that in the adiabatic limit ($\gamma\approx 0$) the slope of $\braket{{p}_r}(p_x)$ is characterized by $\alpha=5/6$. However, on the other hand, taking only the magnetic field after tunneling into account ($\zeta=0$, $\chi=1$) the model predicts a value of $\alpha=-1/6$  which can be shown but does not follow directly from Eq. \ref{Eq4}. Our simulation results show that the value of $\alpha$ does not change by more than 20\% if an ellipticity of $\epsilon=0.8$ is used or spin-orbit splitting or the magnetic quantum number of the electronic ground state is considered \cite{SupplementaryMaterial}. Possible explanations for the deviations comparing the results from the TDSE and the experiment are e.g. multi-electron effects, which are not included in the simulation or the approximation of the atomic potential by a pseudopotential.

In conclusion we have found that the temporospatial structure of the electric field, that has been so far neglected in the literature on tunnel ionization, alters the momenta of the electrons emitted in strong-field ionization and leads to an {\em electric} non-dipole effect. Microscopically the {\em electric} non-dipole effect can be explained by the time-dependent force that acts on the electron in the lab frame. This force is due to the laser field and has a higher frequency for electrons that are traveling anti-parallel to the light propagation-direction than for electrons that travel with the light wave. This change in effective frequency is analogous to the Doppler effect and affects the energy that is transferred to the electron. Thus, the forward-backward symmetry of electron emission in strong-field ionization is broken not only by the well-known {\em magnetic} non-dipole effect but also by an {\em electric} non-dipole effect. We expect that the {\em electric} non-dipole effect will also have an impact on the energetic position of ATI peaks.

\begin{acknowledgments}
\section{Acknowledgments}
A.H. and K.F. acknowledge support by the German Academic Scholarship Foundation. The experimental work was supported by the DFG (German Research Foundation). K.L. acknowledges support by the Alexander von Humboldt Foundation. S.B., M.L. and S.E. acknowledge funding of the DFG through Priority Programme SPP 1840 QUTIF.
\end{acknowledgments} 
\end{document}